\documentclass[sigconf,screen, nonacm]{acmart}
\usepackage{comment}
\usepackage{dirtytalk}
\usepackage{csquotes}
\usepackage[many]{tcolorbox}    	

\definecolor{main}{HTML}{fcdc50}    
\definecolor{sub}{HTML}{fcf2ba}     

\tcbset{
    sharp corners,
    colback = white,
}  

\newtcolorbox{boxH}{
    colback = sub, 
    colframe = main, 
    boxrule = 0pt,
    boxsep = 1pt,
    leftrule = 3pt 
}

\setcopyright{acmlicensed}
\copyrightyear{2025}
\acmYear{2025}
\acmDOI{XXXXXXX.XXXXXXX}
\acmConference[]{}{}{}
\acmBooktitle{,, ,}
\acmISBN{}

\begin{document}

\title{Using Generative AI in Software Design Education: An Experience Report}

\author{Victoria Jackson}
\affiliation{%
  \institution{University of California, Irvine}
  \city{Irvine}
  \country{USA}
  }
\email{vfjackso@uci.edu}
\orcid{0000-0002-6326-931X}

\author{Susannah Liu}
\affiliation{%
  \institution{University of California, Irvine}
  \city{Irvine}
  \country{USA}
 }
\email{susannl5@uci.edu}

\author{Andr\'e van der Hoek}
\affiliation{%
  \institution{University of California, Irvine}
  \city{Irvine}
  \country{USA}
 }
\email{andre@ics.uci.edu}
\orcid{0000-0001-7917-932X}

\renewcommand{\shortauthors}{Jackson, Liu, van der Hoek}

\begin{abstract}
  With the rapid adoption of Generative AI (GenAI) tools, software engineering educators have grappled with how best to incorporate them into the classroom. While some research discusses the use of GenAI in the context of learning to code, there is little research that explores the use of GenAI in the classroom for other areas of software development. This paper provides an experience report on introducing GenAI into an undergraduate software design class. Students were required to use GenAI (in the form of ChatGPT) to help complete a team-based assignment. The data collected consisted of the ChatGPT conversation logs and students' reflections on using ChatGPT for the assignment. Subsequently, qualitative analysis was undertaken on the data. Students identified numerous ways ChatGPT helped them in their design process while recognizing the need to critique the response before incorporating it into their design. At the same time, we identified several key lessons for educators in how to deploy GenAI in a software design class effectively. 
  Based on our experience, we believe students can benefit from using GenAI in software design education as it helps them design and learn about the strengths and weaknesses of GenAI.
\end{abstract}



\maketitle

\section{Introduction}


The advent of freely available off-the-shelf Generative AI (GenAI) tools such as ChatGPT \cite{openaiChatGPT2024} and GitHub Copilot \cite{githubGitHubCopilotYour2024} is having a significant impact on software development. Developers have quickly embraced GenAI tools for a wide range of development activities, including coding \cite{birdTakingFlightCopilot2023}, debugging \cite{kang24_fault}, and generating test cases \cite{wangSoftwareTestingLarge2024}. GenAI has been noted to boost productivity when coding \cite{ziegler_productivity}, improve code quality \cite{mendes_bicycle}, and help identify alternative solutions \cite{khojahCodeGenerationObservational2024}. GenAI tools are only likely to increase across the software profession until they become an integral and indispensable part of software development in much the same way as integrated development environments (IDEs) or continuous integration tools. 

Students, too, have been starting to use GenAI tools for educational purposes. In doing so, some students breach academic integrity rules as they do not declare their work as GenAI generated or assisted, and detectors are not wholly reliable \cite{perkins2024simple}. Rather than prohibiting GenAI use, we feel it is essential to recognize their use is inevitable given their ubiquity. Moreover, we feel learning to use GenAI is essential in preparing students for working in the software profession \cite{acmieeesecurriculum}. Our view aligns with many educators who are also determining the best way to incorporate GenAI tools into their curriculum and for what purpose \cite{kirova_se_education_24}. Some examples include using GenAI to provide automated code feedback to students \cite{nguyenUsingGPT4Provide2024}, explaining code \cite{liuTeachingCS50AI2024a}, and helping students learn code linting \cite{Alomar_softwarequality}.


One key aspect of software development where GenAI has received limited attention is design, an integral part of the software engineering curriculum \cite{acmieeesecurriculum}. Software design helps set the foundation for subsequent development activity by clarifying requirements and, in producing the design, encourages collaboration amongst team members to provide a shared understanding of the system to be built. Moreover, design helps mitigate risks and supports long-term maintenance. 

It is perhaps unsurprising that educators have not yet incorporated GenAI into design education, given there is little research on how professional software designers use GenAI \cite{sengulSoftwareEngineeringEducation2024} for educators to draw upon. To date, the majority of studies of professionals focus on the use of GenAI for modeling activities (e.g., \cite{camaraAssessmentGenerativeAI2023, chenAutomatedDomainModeling2023, jahicStatePracticeLLMs2024}). The handful of studies on software design education studies consider modeling only (e.g., \cite{camaraGenerativeAISoftware2024}) finding that using GenAI to help students learn modeling improves their performance in formal assessments compared to peers who do not use GenAI \cite{camaraGenerativeAISoftware2024}. Benefits of GenAI to designers in areas such as product design include ideation as GenAI quickly provides ideas and alternative perspectives \cite{tholanderDesignIdeationAI2023}, a finding echoed in User Experience (UX) research \cite{takaffoliGenerativeAIUser2024}.

To contribute to the ongoing discussion of how GenAI can be brought into undergraduate software design courses, we provide an experience report where 179 students grouped into 36 teams used GenAI to help complete a design studio assignment requiring the production of a UML class model and supporting pseudocode. In reflecting on their use of GenAI, students felt that: (i) while GenAI helped them in different aspects of their design process, such as generating ideas or validating their proposed solution, its responses required critique and amendments before incorporating them into the final design, and (ii) design is a human-centered activity as humans are essential for driving the process. 
Additionally, in studying how students used GenAI, we noted that the design practices taught in class helped them anchor their GenAI usage. Overall, we feel our experience is a positive one, and the lessons and reflection contained in this paper can hopefully aid other educators in incorporating GenAI into software design classes.


\section{Related Work} \label{related_work}

\subsection{Use Of GenAI in Software Design}

While Large Language Models, and GenAI more broadly, have been well studied in recent years \cite{fanLargeLanguageModels2023}, with many studies focused on coding \cite{houLargeLanguageModels2024}, there is little research that considers how GenAI can aid software design \cite{nguyen-ducGenerativeArtificialIntelligence2023}. One area considered is the use of GenAI to generate the UML models \cite{omgWelcomeUMLWeb2024} commonly used by designers to express their designs \cite{6606618}. 
Studies have found that while, ChatGPT generated UML models were generally syntactically correct, they were not always semantically correct \cite{camaraAssessmentGenerativeAI2023}, missed elements such as relationships \cite{chenAutomatedDomainModeling2023}, and required the diagrams to be manually checked for correctness and completeness \cite{ferrariModelGenerationLLMs2024}. Although a novel ChatGPT-based tool shows promise for improving the quality of use cases \cite{devitoECHOApproachEnhance2023} with designers outperforming those who did not use the tool. Mixed opinions on the value of ChatGPT-generated architectures were also expressed in a study \cite{jahicStatePracticeLLMs2024} examining its capabilities to generate C4 architecture diagrams \cite{c42024}. While its ability to quickly generate C4 diagrams helped provide novel perspectives when ideating the architecture, there were concerns about its need to understand the context and its inability to reproduce its results.



\subsection{GenAI in Design More Broadly}
Beyond SE, design is essential in areas such as user experience (UX) and product design. UX designers perceive that GenAI is helpful for ideation and at the beginning stages of design \cite{takaffoliGenerativeAIUser2024}, and helps to automate monotonous work \cite{liUserExperienceDesign2024}. However, concerns exist about the accuracy of GenAI-generated content due to GenAI not fully understanding the context of the design \cite{takaffoliGenerativeAIUser2024} or its lack of empathy \cite{liUserExperienceDesign2024}. Thus, humans must remain in charge to check the results and to provide empathy and user involvement \cite{liUserExperienceDesign2024, takaffoliGenerativeAIUser2024}. Within product design, studies have also noted how GenAI can aid ideation processes (e.g., \cite{ caiDesignAIDUsingGenerative2023}) by generating ideas quickly \cite{tholanderDesignIdeationAI2023} and augmenting human creativity \cite{chiouDesigningAIExploration2023}. 



\subsection{Use of GenAI in Software Engineering Education}

Software engineering (SE) educators recognize that software engineering education needs to evolve to incorporate GenAI \cite{kirova_se_education_24}. Proposals to change course content include the teaching of higher-level conceptual understanding \cite{kirova_se_education_24}, prompt engineering to encourage helpful responses \cite{dennyPromptProblemsNew2024}, how to critically review and validate the responses to prompts \cite{kirova_se_education_24}, along with a focus on ethics \cite{menzies_ethics}.


Various ways in which GenAI can assist students in their learning include providing personalized feedback to students \cite{liuTeachingCS50AI2024a}, including when coding \cite{nguyenUsingGPT4Provide2024}, and interacting with students via a Socratic approach \cite{al-hossamiCanLanguageModels2024} to help them solve problems. Notably, much research on GenAI in software engineering education has focused on using GenAI to aid students in learning to program \cite{cambazUseAIdrivenCode2024}. Benefits include improvements to computational thinking skills \cite{beckerProgrammingHardLeast2023}, programming self-efficacy \cite{yilmazEffectGenerativeArtificial2023}, and increased performance in code-authoring tasks compared to students who did not use GenAI \cite{kazemitabaarStudyingEffectAI2023}. While recognizing the benefits, educators have raised concerns about the use of GenAI in the classroom including an over-reliance on the use of GenAI, and whether it is appropriate for beginners to use GenAI \cite{cambazUseAIdrivenCode2024} with some evidence that learners with some programming experience may benefit more from GenAI for code generation tasks \cite{kazemitabaarStudyingEffectAI2023}. SE students, too, have concerns about adopting GenAI  including a lack of trust in its responses requiring, a human review  \cite{amoozadehTrustGenerativeAI2024a}. Moreover, a reliance on GenAI to do their work could reduce their own learning \cite{rogersAttitudesUseMisuse2024}.

Little research exists about using GenAI to aid students in learning software design. One paper \cite{camaraGenerativeAISoftware2024} notes that students who used ChatGPT to produce UML class diagrams in formative assessments performed better in the final summative assessment than those who did not use ChatGPT. Students also observed that, while effort was required to use ChatGPT to generate a complete solution, it helped provide a preliminary solution to be manually improved upon later.

\subsection{Software Design Education}
Helping students learn modeling (e.g., UML) is one focus area. Two studies consider how modeling tools can support students in learning software modeling. One focuses on SOCIO, a novel text-based chatbot that supports collaborative modeling. It helped teams of students produce a UML model quicker than those who used an online collaborative modeling tool, although the models were deemed less complete \cite{ren_socio_2023}. In the second, educators note the current suite of modeling tools available is complex and distracts students from learning, so proposed requirements for modeling tools geared towards teaching \cite{kienzleRequirementsModellingTools2024}. 

Some work takes a step back, questioning how to best deliver software design education. The philosophy and pedagogical approaches to designing two new software design courses forming part of a new major degree was shared in \cite{baker_expreport_2009}. Both courses were designed to encourage design-minded thinking and to balance theory and practice. Much of the practice incorporated a design studio approach that emphasizes hands-on in-class activities to aid students learn aspects of software engineering, often using reflective techniques \cite{bullSupportingReflectivePractice2014}. To support student teams who do not have access to a shared, physical space, online whiteboards can aid distributed students to collaborate effectively in design studios \cite{loInfluenceChatGPTStudent2024}.

In summary, little research in both software engineering practice and education explores using GenAI for software design. This experience report adds to this small body of literature on software design education and GenAI. Specifically, our study explores using GenAI to help solve a design problem. In doing so, the students must consider many design aspects, such as identifying the stakeholders, working with constraints and assumptions, and alternative solutions before producing a UML model and pseudocode.



\section{Setting} \label{setting}

\begin{table}
\caption{Expert Practices Taught Prior To The GenAI Assignment. The Observed column notes whether students seemed to apply the practice in the GenAI required assignment. Key: Y = Practice was applied, - = no evidence or cannot tell, N/A = not applicable to the assignment.}
 \begin{tabular}
{p{0.78\linewidth}p{0.14\linewidth}}
\toprule
\textbf{Expert Practice}                                       & \textbf{Observed} \\
 \midrule
Focus on the essence                                            &             Y                           \\
Address knowledge deficiencies                                  &             Y                           \\
Prefer solutions that they know work                            &             -                           \\
Generate alternatives                                           &             Y                           \\
Know design is not done until the code is delivered and running &             N/A                          \\
Dialog between problem and solution                             &             Y                           \\
Solve simpler problems first                                    &             Y                           \\
Draw the problem as much as they draw the solution              &             -                           \\
Go as deep as needed                                            &             -                           \\
Simulate continually                                            &             Y                           \\
Think about what they are not designing                         &             -                    
\\
\bottomrule      
\end{tabular}
\label{table:expert_practices}
\end{table}

This section describes the approach for using GenAI to help students complete their design assignments, contextual information about the software design class, details about the assignments requiring GenAI, and the data analysis performed.

\subsection{Approach} \label{approach}
As noted earlier, little research discusses how GenAI can help students learn to design software. Rather than run an experiment to test what differences GenAI may make to learning software design, we decided to integrate GenAI into the classroom and take a more exploratory approach, whereby students would be required to use GenAI for a single assignment and reflect on their experience. We chose this approach because it allows a safe place for students to learn about using GenAI in an aspect of software development (design) and what benefits it may provide. Given the high adoption rates of GenAI in the software profession \cite{githubcopilotgrowth}, it is important for students to learn about GenAI before entering the workplace. 

The students were asked to use GenAI partway through the course rather than at the start. This delayed introduction would allow the students to receive some instruction on foundational expert design practices (shown in Table \ref{table:expert_practices} and discussed further below) and practice design without GenAI. Introducing GenAI early in the class could lead to a fixation on using it rather than learning how to design.

Quantitatively assessing whether the use of GenAI made a positive difference to the students' overall learning experience did not form part of the remit of this exercise. The exercise was intended to form an understanding of how to integrate GenAI in a meaningful way so students do not simply use it but learn how to use it and its strengths and limitations.

\subsection{Course Overview}
The Software Design course is a 10-week undergraduate course taught at a public University. Each week, there are two lectures (each 80 minutes) and a mandatory discussion section where students can seek help from the teaching assistants while working on the course assignments. Before enrollment, students must have passed three pre-requisite programming-related classes, though no prior design experience is needed to enroll in the Software Design course. As such, most of those enrolled are juniors and seniors, though occasionally, some sophomores who meet the prerequisites also enroll. It is one of six core requirements for the Informatics B.S. degree alongside other courses such as Requirements Analysis and Engineering and Human-Computer Interaction (HCI). Its official course description is, \say{Introduction to application design: designing the overall functionality of a software application. Topics include general design theory, software design theory, and software architecture. Includes practice in designing and case studies of existing designs.} 

The course is purposefully structured to teach students theory and software design practices. 
The course makes extensive use of expert practices \cite{petreSoftwareDesignDecoded2016} used by professional designers to help students learn about the strategies beneficial to software design, along with other design techniques such as UML modeling \cite{omgWelcomeUMLWeb2024}. Amongst the expert design practices taught are focusing on the essence of design problems, generating numerous alternatives to solving a problem, and reusing partial solutions from elsewhere as applicable. The course, too, introduces design at varying levels of detail: (i) designing the functionality of novel applications, (ii) envisioning accompanying user interfaces, (iii) designing high-level architectural solutions, and (iv) specifying detailed UML structures. A theme is that design at these four levels should stay aligned to produce one holistic design that solves the problem. To aid in understanding the four levels, the course touches upon topics such as product visioning and ideation, satisfying stakeholders, and user interfaces relevant to more high-level design alongside topics such as separation of concerns for the more low-level concepts such as modeling software classes.

Collaboration is also a key learning point, so the course is structured such that topics are introduced in a lecture before students work in small teams to practice the topic within the lecture (e.g., identifying the stakeholders for an example design problem, produce a UML class model representing the essence of a problem) and in completing assignments. The class is designed to be a practical introduction to software design to help prepare students for a career in industry designing and building software applications.


In keeping with the practical ethos of the course, the coursework consists of three design studio assignments, each worth 20\% of the final grade. Each design studio asks the students to work in small teams of five to solve a design problem, thus allowing students to practice the design techniques learned in class. The first two design studios comprised two parts, with the second part extending the first. The third design studio was one part only. Each part was graded; thus, five submissions were required in total. Students had three weeks to complete all parts of each design studio.
Regarding GenAI usage in the assignments, GenAI was not mentioned in the assignment details of Design Studio. Its use was mandatory for the first part of Design Studio 2 but optional in Part 2, and its usage was optional for Design Studio 3. Marks were given for using GenAI only in Design Studio - Part 1 (see below).


In total, 179 students took the course in Fall 2023, when this study was set, grouped into 36 teams for each design studio. 

\subsection{Assignments Incorporating GenAI}
In this section, we describe the assignments related to Design Studio 2 and Design Studio 3, as these both allowed the use of GenAI to complete the assignment—-Design Studio 1 did not mention GenAI.

\textbf{Design Studio 2 - Part 1} asked the students to design an educational traffic flow simulator to help civil engineering students understand traffic signal timing. The provided requirements described how students would use the simulator to create a visual map of the area, adjust the behavior of the traffic lights at intersections, and control the flow of traffic, including the density. Simplifying constraints were provided, such as all cars traveling at the same speed and no pedestrian crossings. The assignment required the completion of three design artifacts: (i) a list of stakeholders, (ii) a supporting data structure in the form of a UML Class diagram, and (iii) an algorithm in pseudocode that advances the state of the simulator \say{one tick}. The assignment asked the students to use the expert practices introduced in class, such as solving simpler problems first, generating alternatives (thus encouraging ideation), and focusing on the essence (see Table \ref{table:expert_practices}) to help them solve the design problem. The students were mandated to use GenAI (e.g., ChatGPT) to generate aspects of the design by engaging in at least ten attempts at a conversation with GenAI, and each conversation should be different in purpose. Ten attempts were requested to encourage students to explore alternative perspectives or solve different parts of the problem. At least two team members were requested to collaboratively write each prompt, review the response, and participate in the GenAI conversation. The assignment required the students to submit: (i) a design document containing their identified stakeholders, UML Class diagram, and pseudocode, (ii) the GenAI conversation logs, and (iii) a written reflection on their experiences using GenAI for the design studio. Students had two weeks to complete this first part of Design Studio 2.

\textbf{Design Studio 2 - Part 2} asked the students to extend their initial design to incorporate changed requirements. They had one week for this part and were required to submit an amended design document with the revised UML Class diagram and pseudocode simulating the state of the traffic simulator. In\textbf{ Design Studio 3}, they were asked to solve a problem related to designing an environmentally-themed Augmented Reality (AR) application. The design deliverables for this assignment were higher-level than the previous design studios, thus requiring the students to consider the broader context of how their application would be used and what the users would need. The three deliverables were: the application design (the application features and supported users), interaction design (screen mock-ups of how data would get into the system and how it is presented to users), and architecture design (the main components and the interactions between these components). In both Design Studio 2 - Part 2 and Design Studio 3, using GenAI was optional, although students were asked to provide their conversation logs if they did use GenAI.

\subsection{Training Students on Using GenAI} \label{training}

\begin{figure}
    \centering
    \includegraphics[width=1\linewidth]{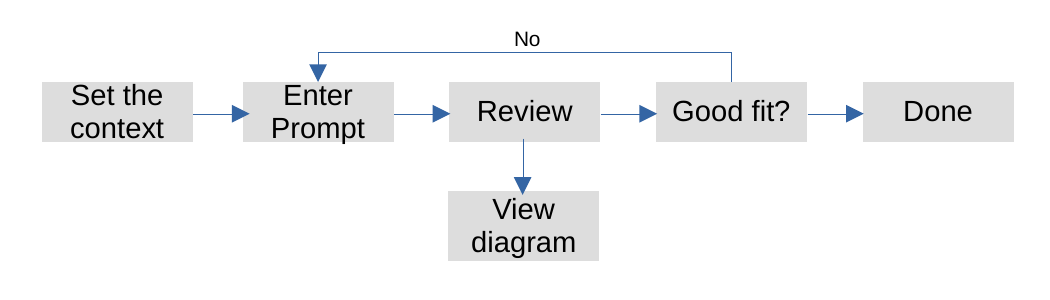}
    \caption{Guidance for generating a design artifact in GenAI}
     \Description{Process flow illustrating the sequential steps to take when using GenAI to generate a design artifact. The context needs to be provided before entering a prompt, reviewing the response (and viewing the result inside a diagram tool), deciding whether the response is acceptable, if not then the prompt should be amended and reviewed again, otherwise the artifact is good enough to be used.}
    \label{fig:llm_training}
\end{figure}
The heart of Design Studio 2 required the students to produce a UML class diagram and use GenAI more broadly to assist in all parts of their design (stakeholders, UML class diagram, and pseudocode). Before the assignment, the students were first introduced to UML class diagrams in a lecture. Among other topics, they were shown how to produce a starting point for their UML class diagram by identifying key pieces of information in a design prompt, such as nouns and verbs. Principles such as separation of concerns, KISS (Keep It Simple, Stupid), and YAGNI (You Ain't Going to Need It) were also introduced to help the students focus on the most critical part of the model. Students were given time in class to practice generating UML diagrams, with the instructor reflecting on their attempts.


Subsequently,
one of the authors used a lecture to demonstrate the use of a GenAI tool (specifically OpenAI ChatGPT 3.5, as this was the most popular text-based GenAI tool at the time of the course and free to use) for assisting in software design, specifically in generating a UML class diagram and pseudocode. Such training was provided as the instructor was concerned that the students had insufficient experience using GenAI, which may lead to a struggle to use GenAI meaningfully or to feel undue stress in using it as part of the assignment. The demonstration was purposefully designed to show how GenAI could be used, but not to be prescriptive in how it should be used, as one of the goals of introducing GenAI in the course was to see how the students would use it creatively. The demonstration emphasized that the approaches shown were just possible approaches. Students were encouraged to experiment with alternative prompt approaches.

The demonstration provided insights into an iterative process (Figure \ref{fig:llm_training}) for generating design artifacts, three high-level prompt engineering strategies, and the usage of ChatGPT in tandem with the publicly available and open-source PlantUML server \cite{plantumlteamPlantUMLWebServer2024} to generate a UML class diagram. PlantUML was selected for use since it generates UML diagrams from text using a simple language, is well-used in software education \cite{agnerModelDrivenEducation2017}, and is supported by ChatGPT \cite{camaraGenerativeAISoftware2024}.

The suggested iterative process of having a conversation with ChatGPT (Figure \ref{fig:llm_training}) emphasized the need to first provide some context before iterating through sequential steps of prompting, reviewing the response, and entering a further prompt to help guide ChatGPT into producing an acceptable design artifact. An example of such context is, \say{You are an expert software architect. I need your help in designing a new application.}

The demonstration highlighted three prompt engineering strategies. Each strategy provided a different way of guiding ChatGPT to a UML model:
\begin{itemize}
    \item \textbf{Brute Force:} Copying/pasting the entire design problem into a prompt and asking ChatGPT to generate a class model, then subsequently reviewing and refining the model by adding classes, relationships, etc.
    \item \textbf{Start small and grow:} Provide a snippet of the problem to generate a small class model before providing further information such as the requirements to refine the model
    \item \textbf{Focus on the problem's goal:} Supply information about the overarching goal of the problem and, through a series of prompts, guide ChatGPT to the problem under consideration.
\end{itemize}

Moreover, the demonstration focused on using ChatGPT to generate a semantically correct UML model, i.e., to create a model containing entities, attributes, and methods meaningfully representing the problem at hand. Students were shown how to generate the UML before using ChatGPT to generate supporting pseudocode.

The design problem used to illustrate the use of ChatGPT for generating a UML class diagram and pseudocode was the same one used in the earlier class period when UML class diagrams were first introduced. Using the same problem allowed students to readily compare their previous models with the GenAI-produced ones so as to begin to understand its strengths and limitations.

\subsection{Data Analysis}
Before the analysis, the assignments were anonymized by removing the student names from the submitted assignments and assigning a random number to each team, hereafter identified as Tx where x is a random number between 100 and 200. Two researchers subsequently analyzed the data. Of the 36 teams, all 36 provided conversation logs, and 30 provided reflections on using GenAI.

The analysis discussed below was only undertaken on Design Studio 2 - Part 1 as few teams provided conversation logs on later assignments (n=15 teams for Design Studio 2 - Part 2, n=14 teams for Design Studio 3) when GenAI use was optional.

Firstly, thematic analysis \cite{clarke_ta} was performed on the reflections provided (n=30). One researcher read each reflection carefully and authored a short memo highlighting interesting points contained in the reflection. These interesting points were subsequently placed onto a FigJam \cite{figjam} board. Once all points were on the board, they were iteratively organized into themes. 

Secondly, one researcher analyzed the prompts and responses in the conversation logs. In total, 360 conversations were analyzed as 36 teams provided the logs, each containing 10 conversations. The analysis focused on the conversations to determine the strategies used by teams when conversing with GenAI, the amount of context put into the prompts, and the types of questions asked. The analysis from each log was cross-referenced with the reflection and design artifacts to provide further context for interpreting the conversation. The observations about GenAI use were reviewed and discussed with the research team before clustering on the same FigJam board containing the themes from the reflections. 

Finally, in reviewing the collated observations, it was noted that some of the expert practices taught in class were visible in the students' reflections and conversation logs. For example, many students talked about \say{essence} or \say{complexity} in their reflections, or some students were asking questions of GenAI that implied they were \say{seeking knowledge}. The data was further reviewed to determine if there was evidence of the expert practices being used. If there was, it was noted and shown in Table \ref{table:expert_practices}.

\subsection{Limitations}
We discuss several limitations in our experimental use of GenAI in our course below.

The students used their own personal ChatGPT accounts, inaccessible to the teaching team. Therefore, we are reliant on the students providing an accurate record of their conversations with ChatGPT. We feel the data provided for Design Studio 2 - Part 1 is likely accurate, given they were required to submit the conversation logs as part of the assignment rubric.

All students used ChatGPT with the majority using the free tier, which at the time of the class was GPT-3.5 Turbo, whereas one team used GPT-4 Turbo, the paid-for tier. The different versions have varying capabilities, leading to different responses for similar prompts. These behavioral differences could impact the students' use and perspectives on the value of incorporating GenAI into software design activities. 
As there was only one team from the 36 that used GPT-4 Turbo, we do not believe the differences in ChatGPT versions significantly influenced our findings. 

We do not have data to investigate why the number of teams that used GenAI dropped when GenAI use was optional. As such, we can only speculate on the reasons behind this in Section \ref{lessons_learned}.

\section{Observations} \label{observations}
This section contains observations about how the students used GenAI in completing the Design Studio - Part 1 assignment. We cover:

\begin{itemize}
    \item How the students engaged with GenAI and for what purpose, e.g., ideation. This includes how they incorporated it into their teamwork.
    \item How the responses contributed to their final design.
    \item Issues encountered when using GenAI.
    \item Student reflections on using GenAI\footnote{As all the teams used ChatGPT, we will use the term ChatGPT rather than GenAI in the remainder of this section} for software design.
\end{itemize}

\subsection{Engaging with ChatGPT} \label{engaging}
Students used ChatGPT in numerous ways including generating ideas, or to seek knowledge, or to check their designs. We also gained insights into how ChatGPT acted as a team-member. We describe these observations further.



\subsubsection{Selective Use} It appears that teams are selective in their use of ChatGPT and chose to only use ChatGPT for a subset of the design deliverables (stakeholders, UML class diagram, algorithm). The majority of teams used ChatGPT to aid with the UML class diagram and the algorithm, with fewer using it to explore the stakeholders. It is unclear why this is the case, perhaps identifying stakeholders was considered an easier task than the others and so assistance from ChatGPT was not required. Nonetheless, across all of the teams, ChatGPT assisted in all three of the design deliverables, showing it has the capabilities to assist in all parts of the asked-for design.

\subsubsection{Ideation}
Seeking alternatives was an expert practice discussed in class so unsurprisingly, many of the teams used ChatGPT to generate ideas for a design artifact. For example, by generating a list of stakeholders. Indeed the ability of the ChatGPT to generate ideas was commonly praised by the teams:

\blockquote{
{\say{\textit{We believe that the place that AI really shines is getting initial ideas and the brainstorming phase of design. This is because ChatGPT has so much knowledge and data to pull from, that when asking it a broad question, it can return many potential answers and ideas. This allows our group to explore many different ideas before deciding on the best one to use for our specific design.}} (T155)}
}


Effective ideation relies upon generating a large set of diverse ideas. Teams benefited from the ideas generated by ChatGPT as they noted it was able to \say{\textit{Quickly generate a range of ideas}} (T130), and that ChatGPT provided a \say{\textit{Diverse range of perspectives}} (T141) and \say{\textit{Alternate approaches and methods}} (T150). Moreover, it was helpful in identifying potential edgecases that the team had not considered.

The teams had mixed opinions on the novelty of generated ideas. Sometimes ChatGPT proposed novel ideas that the team had not thought of, \say{\textit{We ended up using some of the ideas it suggested that we had not thought of.}} (T150) or other times it generated ideas that were the same as ones already identified by the team. This perhaps reflected the varying approaches for when to engage ChatGPT in brainstorming: either first ChatGPT contributes ideas and then the team builds upon them or the team first generates some ideas and then asks GenAI to contribute. 














Although students used ChatGPT to generate ideas, we found little evidence in the conversation logs of students taking advantage of the non-deterministic nature of ChatGPT by asking for further ideas, perhaps by simply repeating the prompt or providing alternative phrasing. The few that did generate alternatives either repeated much of one of the ten conversations in another or asked ChatGPT to generate a design and then immediately asked for an alternative.

\subsubsection{Decision Making} Once alternatives are generated, a decision must be taken as to how best to proceed. This may involve making trade-offs. Some teams used ChatGPT to aid their decision making. Two approaches varying by ownership were observed. In one approach (shown below), the team asked ChatGPT for various suggestions before asking ChatGPT to take the decision as to which one to select, effectively outsourcing the choice to ChatGPT. Notably, ChatGPT always refused to make the decision, instead providing a list of advantages and disadvantages about the various options to better inform the team about potential trade-offs.
\begin{boxH}
\textbf{T177:} \say{You are an expert software designer tasked with creating a traffic flow simulation
program for students to use in a civil engineering class. Give 5 different ways the program can decide which direction each car will go (straight, left, or right)?}\\
\textbf{ChatGPT:} \textit{Provides 5 suggestions}\\
\textbf{T177:} \say{Would you have the user choose a destination for each car or have the program choose?}
\end{boxH}

In the second, the team made the decision aided by ChatGPT as it provided an alternative viewpoint:
\blockquote{
\say{\textit{When we were considering two options we realized that we could ask ChatGPT about the situation we were trying to solve and ask it for its solution, comparing it to ours. This leads to another valuable opinion and helps the group overcome hurdles when we were stuck on how and where to proceed.}} (T155)}

\subsubsection{Knowledge Seeking}
Another expert practice discussed in class was the need to \say{Address knowledge deficiencies} when designing. This practice refers to the need for designers to identify any knowledge gaps that may hinder their understanding of the design problem and to address the gaps as early as they can. As such, we noted some teams used ChatGPT to seek knowledge to aid their design process. For example, one team asked how sensors at traffic lights work.

\subsubsection{Validation of Design}
Some teams used ChatGPT to validate their designs by using it as a \say{\textit{Second-pair of eyes.}} (T146) That is, they generated the design themselves, provided this design along with information about the design problem, and asked ChatGPT to validate the design against the problem.

\begin{boxH}
\textbf{T112:} \say{For the prompt below, we currently have the classes for the UML Diagram, which are
[Listed 12 classes] Is there any other classes we are forgetting? Here is the prompt: [Pasted in the entire design prompt]} \\
\textbf{ChatGPT:} \textit{Provides 9 additional classes, each with a description} 
\end{boxH}

Other teams also used ChatGPT to check consistency between two different artifacts:

\blockquote{
\say{\textit{We asked the AI to compare our UML diagram and pseudocode to the prompt to check if our solution may have been missing any components. We also asked it to evaluate the UML diagram and pseudocode with each other to ensure that the two were compatible.}} (T176)}

Using GenAI to check their designs has the potential to help teams to improve their design quality both in terms of: (i) ensuring requirements were satisfied as GenAI was able to identify omissions, and (ii) consistency by recommending naming conventions, meaningful function, and variable names. 

\subsubsection{Other uses}
There were some unexpected uses of GenAI in the assignment. One team used ChatGPT to test the accuracy of their UML class model by asking it to simulate the simulation and to identify any missing methods. In doing so, ChatGPT identified missing methods which the team subsequently asked ChatGPT to incorporate.

Reflection is an important aspect of design \cite{petreSoftwareDesignDecoded2016}. Some teams asked ChatGPT to reflect on the design it generated, as shown below. This is also another case where the team delegates the decision about what to change to ChatGPT.

\begin{boxH}
Team had previously asked ChatGPT to generate the pseudocode\\
\textbf{T183:} \say{Can you talk about the weakness of your algorithm?}\\
\textbf{ChatGPT:} \textit{Provides weaknesses} \\
\textbf{T183:} \say{And also tell me about some of the good points of the algorithm as well} \\
\textbf{ChatGPT:} \textit{Provides some good points} \\
\textbf{T183:} \say{As considering the weakness of the algorithm, can you make more improvements?} \\
\textbf{ChatGPT:} \textit{Provides revised pseudocode}
\end{boxH}

\subsubsection{Collaboration}
In many ways, ChatGPT acted as a valued team-member able to support and coach the team. We noted that it supported the team in various ways, including helping to spark new ideas as the team riffed off the ideas provided by GenAI, supplied alternative perspectives to those identified by the team, unblocked the team when stuck, or \say{\textit{Jumpstart the sessions when we reached lulls or obstacles}} (T183), or even providing a \say{\textit{Platform to bounce ideas off.}} (T153) By reviewing its responses, it can also help the team better understand the problem, \say{\textit{Including another perspective in our brainstorming process helped to identify other edge cases and further our understanding of the problem at hand.}} (T130). This latter point perhaps shows how ChatGPT can help teams consider both the problem and the solution space when designing, an important consideration for any designer \cite{dorst2001creativity}.

One team mentioned that because it was viewed as being \say{\textit{An impartial platform for idea growth and confirmation}} (T122), it can improve team dynamics. This is an interesting perspective given students' teamwork struggles when working on team projects \cite{donelan2024online}. 
 
Another team found it \say{\textit{Incredibly helpful}} 
 (T122) that ChatGPT could generate PlantUML code that could be readily visualized. The visual model helped them to understand and communicate the system's architecture making it easier for the team to collaborate.







\subsubsection{Prompt Strategies}
We saw the use of all three prompt strategies described in the training (Section \ref{training}). However, some teams used more novel approaches. For space reasons, we provide one example only: 

\begin{boxH}
\textbf{T143:} Team provided the context to the design problem and ends with \say{Ask any clarifying questions if you have any. Otherwise, say: I understand. You will then be given the requirements if you have no questions.}\\
\textbf{ChatGPT:} \textit{Asks for more specific requirements} \\
\textbf{T143:} Provides the detailed requirements and ends with \say{What meaningful questions do you have} \\
\textbf{ChatGPT:} \textit{Asks a number of questions} \\
\textbf{T143:} Answers each question and ends with \say{Do you have any more questions? If you do, ask them. Otherwise, if you do not have any more questions, say: I am ready.} \\
\textbf{ChatGPT:} \textit{Has no more questions and says it’s ready} \\
\textbf{T143:} \say{Okay. You will first begin with a UML design for the program including the data structures and methods. Keep in mind the algorithm to advance the simulation state. Avoid meaningless data structures or methods. Return the design as PlantUML}
\end{boxH}

This strategy seems carefully considered to ensure that ChatGPT has sufficient information to generate a meaningful response. Whether it is more successful than the simpler approaches shown in the training is unknown, but it shows how certain teams clearly have significant experience in interacting with ChatGPT in advanced ways.

We also observed variance in the length of conversations. Some teams engaged in lengthy back-and-forth conversations whereas others were very terse with only one or two prompts. The lengthy ones sometimes were due to the team iterating with ChatGPT to revise a model by drip-feeding the requirements, offering decisions made by the team, or correcting the model when they thought it had extraneous or missing data. 

\subsection{Incorporating ChatGPT Responses into Design}
Teams overwhelmingly noted that ChatGPT provided a design that was a helpful starting point. Yet they often spotted issues with ChatGPT's designs resulting in user-initiated amendments either through ChatGPT or in a separate design tool.

\subsubsection{Provides a Starting Point}
ChatGPT often gave a good starting point for the team's design, be it a list of stakeholders, the UML class diagram, or the pseudocode for the algorithm. Sometimes the team was experiencing difficulties getting started so appreciated that ChatGPT was able to generate a starting point for them to develop further, \say{\textit{It definitely helped us in getting something down on paper right off the bat. This saved us a lot of time since it is often hard to get started in the beginning.}} (T199) 

\subsubsection{Critiquing and Modifying The ChatGPT Generated Design}
Arguably, the underlying reason many teams considered the ChatGPT generated design as merely a starting point was due to inaccuracies that teams spotted in the design (discussed in Section \ref{issues_llm}). These inaccuracies meant the design would not satisfy the requirements in some way, perhaps because it was missing some aspects or it had aspects that were considered out-of-scope and thus unnecessary. Many teams recognized a review was necessary as ChatGPT was not perfect, and that its responses should be taken as suggestions, treated with skepticism, and that \say{\textit{Its suggestions should be critically evaluated.}} (T141) 

For some, realizing ChatGPT was infallible, was a learning point, \say{\textit{At first, I unfortunately made the mistake of taking its responses too seriously; that is, I thought that whatever it cooked up was going to be in the final design. This caused some panic in me because I did not get ChatGPT to give me satisfying results}}(T146).

When deciding what parts of the design to keep, there was evidence of cherry-picking which parts to keep and which to throw away, \say{\textit{Our group had to mull over possible benefits and ramifications of including or excluding suggestions the AI had made.}} (T180) Modifications too were often made within ChatGPT by requesting changes to the pseudo code or adding new classes or relationships to the class diagrams, as in the following example:

\begin{boxH}
Team had already asked ChatGPT to generate the class model\\
\textbf{T101:} \say{What about students as a class?}\\
\textbf{ChatGPT:} \textit{Updates the class model with students} \\
\textbf{T101:} \say{we aren't gonna have any roads that start within the middle of the map. All roads start and end on the edge of the map}\\
\textbf{ChatGPT:} \textit{Updates the class model} \\
\textbf{T101:} \say{I think that road should be separated from the intersection class, intersection can be a child right?}\\
\textbf{ChatGPT:} \textit{Updates the class model} 
\end{boxH}

Other times, teams manually drew their own class diagrams in a different tool such as LucidChart \cite{lucidchart} or Draw.io \cite{drawio}, but often based on what ChatGPT produced, \say{\textit{We moved forward with expanding from what ChatGPT had and what we came up with as well to create our UML diagram.}}(T133)

At times, teams were unhappy with the response, so designed it themselves, \say{\textit{How a tick should be measured within the simulation was hard for the model to understand and implement in a logical manner, so it was designed manually.}} (T143)   

\subsection{Issues With ChatGPTs Responses} \label{issues_llm}
Many teams commented on issues encountered when reviewing ChatGPT's response. These issues ranged from somewhat irksome ones (e.g., syntax errors) to more fundamental ones where the teams questioned the value and usefulness of the response (e.g., undue complexity, superficiality). 

\subsubsection{Syntax Errors}
More an annoyance than anything else, but teams observed syntax errors in generated PlantUML code and also the pseudo-code. This led to additional work for the teams to correct the generated code, \say{\textit{Sometimes it will generate things with index errors and you need to debug the plantUML codes.}} (T112)

\subsubsection{Reliability}
Teams noted that ChatGPT could be unreliable. Sometimes ChatGPT would forget things discussed earlier in the conversation, made inaccurate assumptions instead of asking for additional information, contradicted itself, or left out details. This resulted in additional labor such as reminding it of previously discussed requirements: 

\blockquote{\say{\textit{ChatGPT will sometimes contradict itself or leave out details from earlier in the discussion, as the conversation progresses, so we had to keep double checking its work and reassessing that the work we incorporated from it was correct.}} (T171)
}

One recurring issue in using PlantUML was the struggles ChatGPT had with incorporating relationships. It was unable to distinguish dependency and aggregation relationships, and omitted key relationships between classes. Even when corrected, it was unable to incorporate the change into the PlantUML. This inability to model relationships resulted in additional work for the team to manually update the UML diagram with the desired relationships.

Finally, the behavior of ChatGPT caused frustrations as it would \say{\textit{Not do what it was supposed to do no matter how many times we would ask it.}} (T151)

\subsubsection{Introducing Complexity}
The teams noted that the responses often contained, from their perspectives, unnecessary complexity that they therefore did not incorporate into their design. Two examples of how complexity was proposed by ChatGPT: (i) offering suggestions deemed by the team to be outside the scope of the requirements, (ii) including additional classes such as helper classes that were not needed to address the essence of the problem in UML diagrams. Having to review the designs for such complexity caused an unnecessary distraction for some, \say{\textit{It often began making unnecessary changes or additions that distracted from tackling the real problem.}} (T199)

Teams applied the expert practices (Table \ref{table:expert_practices}) to help them identify potential complexities, specifically by focusing on the essence of the problem and preferring simpler solutions:

\blockquote{\textit{Adding in sound effects, gamification, and a tutorial/help system strays too far from the essence of the software. Plus, the software is meant to be simple, making help and tutorials unnecessary.} (T154)}

\subsubsection{Superficiality of the Responses}
Teams noted that the responses returned from ChatGPT were unhelpful as the responses were too general or lacked depth to really address the prompt. Again, the use of the taught expert practices helped them realize the response was not relevant to their design, in this case focusing on the essence, \say{\textit{The generality of the answers meant that they often strayed from the essence of the problem.}} (T183)

\subsection{Student Reflections on ChatGPT for Software Design}
When reflecting on their use of ChatGPT for the assignment, there was a consensus that it was beneficial for reasons such as giving them a good starting point for a design and boosting their productivity. However, many teams felt it could not be used by itself, and designers still had a critical role to play in shaping the design.

\subsubsection{Helpful For Some Aspects Of The Design Process}
Many students noted that while ChatGPT was helpful, they felt it could only be used for parts of the design process, especially at the start when faced with an empty canvas. They got their design kick-started by ChatGPT's ability to quickly generate many ideas that could be discussed and taken forward into the design. One team noted that getting started, was \say{\textit{Usually the hardest part}} (T100), and so it helped them get over the initial hump faced when beginning any new project. 

Other teams found it helpful for generating initial concepts, broadening their perspective on an issue, or working on smaller design aspects. Teams felt ChatGPT was less beneficial once the design got larger and slightly more complex. None of the teams felt it was good enough to complete the entire design without the team leading the process, \say{\textit{ChatGPT is a good polisher, examiner, and narrator, but never the sole producer.}} (T146)

\subsubsection{The Role Of Human Designers}
Overwhelmingly, the students felt the human designers needed to lead the design rather than being wholly dependent on ChatGPT to generate the design. They felt humans were essential for the critical thinking and decision-making required when designing, as these were tasks ChatGPT could not perform due to the lack of contextual understanding and intuition that human designers have. As noted earlier, students needed to sense-check ChatGPT's suggestions to ensure the design was not becoming overly complicated, straying away from the essence of the problem, or just making sense. Moreover, some students noted that ChatGPT-generated responses lack creativity, so blending human ideas with ChatGPT-generated ideas is essential to ensure the design is creative.

Some felt it was important that designers should rely on their own skills and creativity to generate an initial design before using ChatGPT, as there is a risk that \say{\textit{Designers could potentially limit their understanding of the problem by relying on them too much.}} (T111) 

Another student alluded to a risk that outsourcing too much to ChatGPT could lead to a lack of understanding of the design:

\blockquote{\say{\textit{It is easy to have ChatGPT do the thinking for you, but it is dangerous - knowing why a design decision is pursued is critical for learning, and it is easy to simply copy its response without prying into how it came to fruition.}} (T146)}

\subsubsection{Productivity}
Similar to studies on coding \cite{weberSignificantProductivityGains2024}, students noted a perceived boost to their productivity by using ChatGPT and so were able to complete their designs more quickly. One team even felt its use \say{\textit{Definitely cutting down our whole process in half.}}(T132) The productivity improved for several reasons, including the speed at which ChatGPT generates ideas and the automation of \say{busy} work through its ability to generate structured formats such as PlantUML that could be quickly improved upon through iteration based on feedback from the team.

One team observed that cost-benefit trade-offs had to be made when deciding whether to persist in refining the model via ChatGPT or stop using it and take the design into a tool for manual updates. 

\subsubsection{Helpful Learning Experience}
Some students appreciated the opportunity to use GenAI in their assignments, noting that it was an interesting and eye-opening experience and that it helped to highlight the potential of \say{\textit{AI-assisted design}} (T180), more so as its use is banned in other classes. One team noted that, for all of the team members, this was the first assignment at the University that they were \say{\textit{Granted permission to use AI without being penalized for plagiarism or academic dishonesty.}} (T181)

Perhaps because of this lack of experience, some teams struggled with finding prompts that gave reasonable answers and felt disappointed with ChatGPT. Others, however, used the class as an opportunity to learn how to author prompts and also how to best incorporate ChatGPT into their design process:

\blockquote{
\say{\textit{In the beginning, we simply input the prompt and had it generate the project for us. However, most of the output was overwhelming and for some of the output we didn’t know how to utilize it to best benefit our project. As we progressed throughout the project, we realized ChatGPT was a good way for us in creating starter points for our designs, such as our UML diagram and pseudocode.}} (T151)}

\section{Reflections} \label{reflections}

This experience report on the use of GenAI within a software design education class contributes to the growing body of research on using GenAI for software development education. In our class, the students learned about GenAI and their potential for software design, and, for many, it was the first time they were actively providing training and encouragement to use GenAI in assignments. We provide lessons learned below for educators interested in repeating our approach to introducing GenAI into software design classes. We conclude this section with some thoughts on future work.


\subsection{Lessons Learned} \label{lessons_learned}

\textbf{Lesson Learned \#1:} \textit{GenAI can be a useful tool to aid software design education.}
We learned that ChatGPT can be a useful tool in software design education. It helps students get started, generate ideas, validate their solutions, and speed up their process. Issues are encountered, though, such as its unreliability and tendency to introduce complexity. These issues show it cannot be relied upon to produce a complete design. The students' experiences show that GenAI is not a threat to student designers. It is considered a tool that can supplement human design skills. The students firmly believe that human designers need to continue to own the design process by deciding when and what to incorporate into the design through thoughtful prompting and critical review of the responses. This perspective is similar to findings in other design areas (e.g., \cite{tholanderDesignIdeationAI2023, liUserExperienceDesign2024, takaffoliGenerativeAIUser2024}). Having a human-centered design process aligns with views from HCI scholars such as Shneiderman, who argues that AI systems need to be aligned with human needs and values and they should enhance human capabilities \cite{shneidermanHumanCenteredAI2022}, not replace humans. For educators, we cannot expect or encourage the students to use GenAI for the entire design. Instead, we must teach them that it can help in various ways and that students must acknowledge its limitations and still engage in the design to ensure its validity.

\textbf{Lesson Learned \#2:} \textit{Reflections helpful.}
We feel that asking the students to reflect on their experiences of using GenAI helped them reinforce the advantages and disadvantages of using GenAI for software design. 
Notably, some students demonstrated advanced prompting strategies, presenting an opportunity for knowledge sharing within the class. In addition to asking students for their written reflections, we recommend incorporating the opportunities for open discussion in class. Students showcased creative prompting strategies that went beyond the prompt engineering examples we provided. It would be beneficial to discuss these examples with the entire class and hear from the students their motivations, thoughts on what worked well, and what they did with the result. 

\textbf{Lesson Learned \#3:} \textit{Expert design practices help in anchoring GenAI usage.}
The expert practices we teach in our class (see Table \ref{table:expert_practices}) serve as a crucial framework, guiding students in their use of GenAI in terms of both purpose for using it (e.g., seeking knowledge, generating ideas, alternative designs) as well as critiquing the responses (e.g., focus on the essence and preferring simplicity) and continuing the conversations to steer the GenAI towards these practices. Educators have identified the need to teach critical thinking skills to students to aid them in reviewing GenAI responses \cite{kirova_se_education_24}, and it appears the expert practices can serve as a framework for doing so. Further, we feel our approach of introducing GenAI after teaching (and practicing) some foundational expert practices in lectures and in a first design studio where GenAI was not required and discouraged helped the students contextualize the subsequent use of GenAI. We recommend a similar sequencing to others wishing to introduce GenAI: teach some basic design first so students become sensitive to the topic before introducing GenAI.

\textbf{Lesson Learned \#4:} \textit{Importance of prompt engineering.}
As highlighted in \cite{dennyPromptProblemsNew2024}, prompt engineering is a core practice for students to learn. We provided some training to help students start on prompting, though it was not exhaustive as we were curious how students would engage. It is thus perhaps not surprising that some students struggled to get reasonable responses from GenAI due to inadequate prompts. Moreover, some teams followed the taught approaches rigidly and did not explore alternative prompting strategies. Contrastingly, we saw some highly novel prompts in both the strategy to coax a design from GenAI and in undertaking expert practices such as validating the design either directly or by asking GenAI to self-reflect or to simulate the execution of the pseudocode. This contrast in prompting reflects varying prior experiences of prompt engineering, which is to be expected given the large size of the class. It also illustrates that some students were perhaps more advanced at prompting than the educators since some prompts were unexpected and surprising to the instructor. Further training before using GenAI could help inexperienced students in prompt engineering. Yet, there is a trade-off, as too much training could lead to more copycat usage rather than student-initiated exploration and reflection. We recommend educators consider how much training is required, which could vary depending on the students' prior experience. We also recommend educators consider how to encourage the students to engage in lengthier conversations with GenAI as we observed that only 30\% of the ten conversations from each team consisted of four or more prompts. Rather than asking for 10 conversations, an alternative might be to encourage greater dialog with the LLM but in fewer conversations. For example, a different strategy could be to ask the students to use different prompt engineering strategies for the same purpose to assist in their design.

\textbf{Lessons Learned \#5:} \textit{Unclear why GenAI usage reduced when optional.}
An outstanding puzzle was the reduced use of GenAI when it was optional (less than half the teams appeared to use it) despite the reflections showing that many teams found it beneficial to get them started or to generate ideas. It needs to be clarified if the use did indeed drop off, and if so, why. Perhaps GenAI was not well suited to assist in the design prompt in the third design studio, although this seems unlikely given that some teams continued to use GenAI even when its use was optional. Perhaps students used GenAI but did not submit the logs as there was no incentive (in the form of marks on the assignment) to do so. This puzzle is worth exploring further. We recommend that educators should: (a) provide their own GenAI chatbot that automatically logs the conversations, negating the need for students to provide the logs, and (b) offer some points in the assignments when GenAI use was optional for either reflecting on its use or on reasons why they did not use it.

\textbf{Lessons Learned \#6:} \textit{Societal concerns on GenAI.} None of the reflections mentioned wider societal issues associated with GenAI, such as ethics, copyright, or sustainability concerns typically expressed by professionals \cite{liUserExperienceDesign2024}. It is unclear if the students are unconcerned about such issues or whether the wording of the reflection exercise was insufficient to encourage such philosophical reflection. Educators are encouraged to help students consider the societal impacts of GenAI \cite{kirova_se_education_24}, so we recommend instructors consider: (i) encouraging students to reflect more broadly on the use of GenAI beyond its immediate practical use, and (ii) depending on the wider curriculum in which the design class resides, whether an in-class discussion on societal impact is warranted.

\subsection{Future Work}
We have provided an experience report, but clearly, research is needed to explore some of the observations further. In particular, we did not determine whether GenAI makes a difference in learning outcomes as the focus was on students learning more about its capabilities for supporting design. Accordingly, defining appropriate measures of learning, examining how the use of GenAI affects success in future courses and careers, and novel tooling to assist learning outcomes could all be considered further. Exploring alternative approaches to the one outlined here is also needed. For example, breaking up the single assignment with optional use afterward into smaller, more targeted assignments where each focuses on different aspects (e.g., idea generation, design verification) may lead to improved student learning. Beyond, we foresee two important directions:

\textbf{Research \#1:} \textit{Customized chatbot.} We believe the expert practices provide a strong foundation for students to learn about software design and, for some teams, helped in their usage of GenAI. It could be beneficial to provide a customized chatbot interface around an LLM (e.g., OpenAI APIs) that helps all students apply the expert design practices taught in the class. Such a chatbot could encourage the student to identify where they have knowledge gaps or to generate alternatives before narrowing down to a preferred option. 

\textbf{Research \#2:} \textit{Does the use of GenAI improve teamwork in software engineering education?} Some of the students highlighted that using GenAI improved their teamwork due to its perceived neutral stance and for helping teams better understand the context of the problem. Studying the use of GenAI in student teamwork, in the context of software design or other team-based software engineering classes, could provide further insights into how GenAI aids teamwork.

\section{Conclusion} \label{conclusion}
This paper provides an experience report of introducing GenAI (specifically ChatGPT) into an undergraduate software design class. Students were required to use GenAI in a team-based assignment to produce a design for a stated problem. By analyzing the students' reflections on using GenAI and the conversation logs, we observed that students found GenAI helpful in several ways, including generating ideas, getting them started, improving their teamwork, and reducing the time it took to complete their design. However, they all noted the necessity for humans to lead the design process and not rely entirely on GenAI for reasons such as its inability to understand the context or its tendency to overcomplicate the solution. As such, we believe our approach of introducing GenAI after some fundamental design practices have been taught and asking students to reflect on their experiences has helped students learn about the strengths and weaknesses of GenAI for software design. 
We hope our experience will encourage other educators to adopt GenAI into their software design classes.


\bibliographystyle{ACM-Reference-Format}
\bibliography{main}

\end{document}